# Femtosecond Laser Filamentation in Atmospheric Turbulence

JIEWEI GUO[1,2#], LU SUN[1,2#], YUEZHENG WANG[1,2], JIAYUN XUE[1,2], ZHI ZHANG[1,3*], HAIYI LIU[1,2], SHISHI TAO[1,2], PENGFEI QI[1,2], LIE LIN[1,3], WEIWEI LIU[1,2]

[1] *Institute of Modern Optics, Eye Institute, Nankai University, Tianjin 300350, China*
[2] *Tianjin Key Laboratory of Micro-scale Optical Information Science and Technology, Tianjin 300350, China*
[3] *Tianjin Key Laboratory of Optoelectronic Sensor and Sensing Network Technology, Tianjin 300350, China*
*\* [zhangzhi@nankai.edu.cn](mailto:zhangzhi@nankai.edu.cn)*

**Abstract:** The effects of turbulence intensity and turbulence region on the distribution of femtosecond laser filaments are experimentally elaborated. Through the ultrasonic signals emitted by the filaments, and it is observed that increasing turbulence intensity and expanding turbulence active region cause an increase in the start position of the filament, and a decrease in filament length, which can be well explained by the theoretical calculation. It is also observed that the random perturbation of the air refractive index caused by atmospheric turbulence expanded the spot size of the filament. Additionally, when turbulence intensity reaches $8.37 \times 10^{-12} \, \mathrm{m}^{-2/3}$, multiple filaments are formed. Furthermore, the standard deviation of the transverse displacement of filament is found to be proportional to the square root of turbulent structure constant under the experimental turbulence parameters in this paper. These results contribute to the study of femtosecond laser propagation mechanisms in complex atmospheric turbulence conditions

## 1. Introduction

When the femtosecond laser pulses propagate in transparent media including gases, liquids, and solids, the beam can propagate over a long distance without diffraction along with a self-generated plasma channel named filament.[1-3] The formation of the optical filaments the balancing result of the dynamic counteraction of the optical Kerr effect induced self-focusing and the defocusing [2, 4]. The femtosecond laser filamentation can give rise to a nearly constant laser intensity as high as $10^{14}$ W/cm$^2$ in a long distance ranging from meters to kilometers in the atmosphere[5, 6]. The ultra-strong laser intensity is high enough to induce remote ionization and fragmentation of molecules, giving rise to characteristic fingerprint fluorescence emissions and providing an opportunity to remote sensing for identifying parent molecules in a variety of complicated environments, especially the remote sensing for identifying molecules[7-9]. The length, position, and intensity of the femtosecond laser filament can be effectively controlled by initial laser parameters and the spatiotemporal shaping of the pulses[10-12]. Therefore, the versatile femtosecond laser filament technology demonstrates a great potential for the remote sensing of air pollutants and hazardous matters such as explosives and chemical weapons. However, the remote sensing of multicomponent pollutants and targets remains challenging owing to the inevitable turbulence and ultralow concentration.

Femtosecond laser filaments are inevitably affected by atmospheric turbulence in air, which poses an unavoidable problem for the practical application of femtosecond laser filaments. Recently, some studies have been reported on the effect of atmospheric turbulence on femtosecond laser filament formation[13, 14]. Turbulence induces the fluctuation of the surrounding air density, consequently causing refractive index variations and distorting optical waves, which leads to a transverse stochastic distribution, increased filament counts and expanded spot size [15-17]. Moreover, the onset of filamentation has been observed to be contingent upon different conditions[18-21]. These trivial changes in filamentation have a great impact on sensitivity during remote sensing due to the ultra-strong nonlinear effect. Hence, the spatial distribution of filament during their propagation in turbulent atmosphere plays a critical role in remote sensing. The dynamics of long-range femtosecond laser filamentation in atmospheric turbulence is not clear yet. The empirical formulas describing turbulence effects on the pointing stability of optical filaments are not yet to be reported, which will be also a crucial issue for achieving a high-performance remote air laser.

Here, we focus on the femtosecond laser filamentation in turbulence with a distance of 30 meters. We carefully examined the impact of turbulence intensity and interacted region on the onset distance, length, intensity, and diameter of the filaments. The increased turbulent intensity and interacted region increased the onset and length of filament, which can be well explained by the theory. Furthermore, the empirical formulas describing the relationship between the standard deviation of transverse displacement of filament and turbulence intensity in different regions were proposed.

## 2. Experimental setup

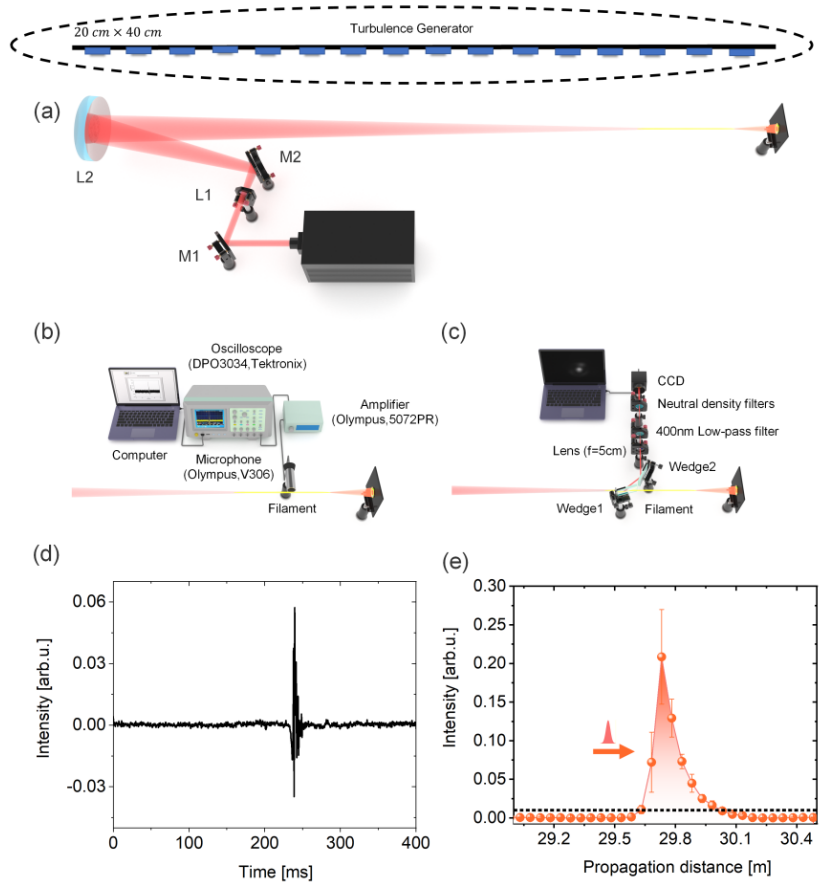

Fig. 1. Schematic diagram. (a) Schematic diagram of the experimental setup; (b) Experimental setup for recording ultrasonic signal induced by filament; (c) Experimental setup for recording the cross section of filaments; (d) A time-domain diagram of the filament-excited ultrasonic signal; (e) Longitudinal acoustic intensity along the filament without turbulent (dashed line is 3σ).

The experimental remote sensing setup using filaments is illustrated in Fig. 1. A commercial femtosecond laser system (Legend Elite, Coherent Inc.) delivers 60 fs pulses at the central wavelength of 800 nm, with the maximum pulse energy of 4 mJ at a repetition rate of 500 Hz. The laser pulse output from the laser system was focused by a lens group comprising of a concave lens(L1) with a focal length of -150 mm and a convex lens(L2) with a focal length of 203 cm. The geometrical focus of the lens group is located 30 m away from the convex lens. The position of L2 is defined as the original point of the propagation distance $z = 0$ m. Along the propagation of the femtosecond laser pulses, air turbulence is introduced by the turbulent blower which consists of a 30 m-long air duct with two high-speed centrifugal turbo blowers (1.2 kW) placed above the laser propagation. By manipulating the airflow outlet below the duct, we have the capability to modify the turbulence region.

To clarify the onset and length of filament, as shown in Fig. 1(b), an ultrasonic probe (V306, Olympus. Ltd.) combined with an amplifier (5072PR, Olympus. Ltd) and an oscilloscope (DPO3034, Tektronix Inc.) is placed at a distance of 1 cm perpendicular to the propagation axis to collect the ultrasonic wave emitted from the optical filament. The typical time domain ultrasonic signal is shown in the insert in Fig. 1(d). Because the length of optical filament (~30 cm) is much longer than the spatial resolution (~0.87 cm) of the ultrasonic microphone, the microphone is mounted on an electrically driven sliding rail and moved parallel to the laser propagation direction to measure the length and distribution of the filament point by point. The spatial step of the microphone is 2.5 cm.

To analyze the distribution of the filament cross section, as indicated in Fig. 1(c), two fused silica wedges are inserted into the laser beam path, both at grazing angles, yielding a reflectivity of about 10% at each front surface. Therefore, after two surface reflections, the laser intensity is reduced to approximately 1%. The cross-sections of the laser beam are then detected by a CCD camera through a calibrated 1:4 image setup. The exposure time of the CCD is set as 1 ms to capture a single pulse for each picture. The neutral density filters are placed in front of the CCD camera to further attenuate the laser intensity.

By adjusting the frequency of the blower and the location of the outlets, we can control the intensity and position of turbulence, respectively. The turbulence intensity is expressed by the refractive index structure constant: $C_n^2 = \sigma^2 \phi^{1/3} / 2.91 L$, where $\sigma$ is the standard deviation of the angle of arrival. The beam diameter and the length of the turbulence region are denoted by $\phi$ and $L$, respectively. To calculate the refractive index structure constant, a low-power continuous-wave He-Ne laser was used, which propagates along the same path as the femtosecond laser pulses. We quantified the intensity of the artificial turbulence in this paper, and the refractive index structure constants are $1.68 \times 10^{-12} \text{m}^{-2/3}$, $3.11 \times 10^{-12} \text{m}^{-2/3}$, $3.83 \times 10^{-12} \text{m}^{-2/3}$, $8.37 \times 10^{-12} \text{m}^{-2/3}$, $1.61 \times 10^{-11} \text{m}^{-2/3}$, $2.5 \times 10^{-11} \text{m}^{-2/3}$ which relatively correspond to atmospheric turbulence[15, 21].

The distribution of ultrasonic signal along the filament without turbulence as a function of the propagation distance z is shown in the Fig. 1(e), where the start and end positions of the filament are defined at the positions where the ultrasonic signal value is just over the dashed line according to the 3σ rule (σ, the standard deviation of the background noise intensity). To reduce the error of experimental measurements, the measurement in Fig. 1(e) is the average result of ten measurements. And the start of the filament is about 29.71 m.

### 3. Results and Discussion

First, the longitudinal distribution of the filament under different turbulent conditions are investigated. Fig. 2(a) depicts the recorded onset distance of filament with different turbulent intensities with turbulent region of 30 m, which can significantly modulate the filamentation dynamics and the acoustic emission. It can be clearly observed that the increased turbulent intensity increases the onset distance of filament. It is worth noting that the minimum turbulence intensity depicted in Fig. 2(a) reflects the air disturbance within the experimental environment. Furthermore, it is observed that altering the region of turbulent activity does not modify the correlation between the location of filament initiation and the intensity of turbulence (Fig. S1 in Supplementary Materials).

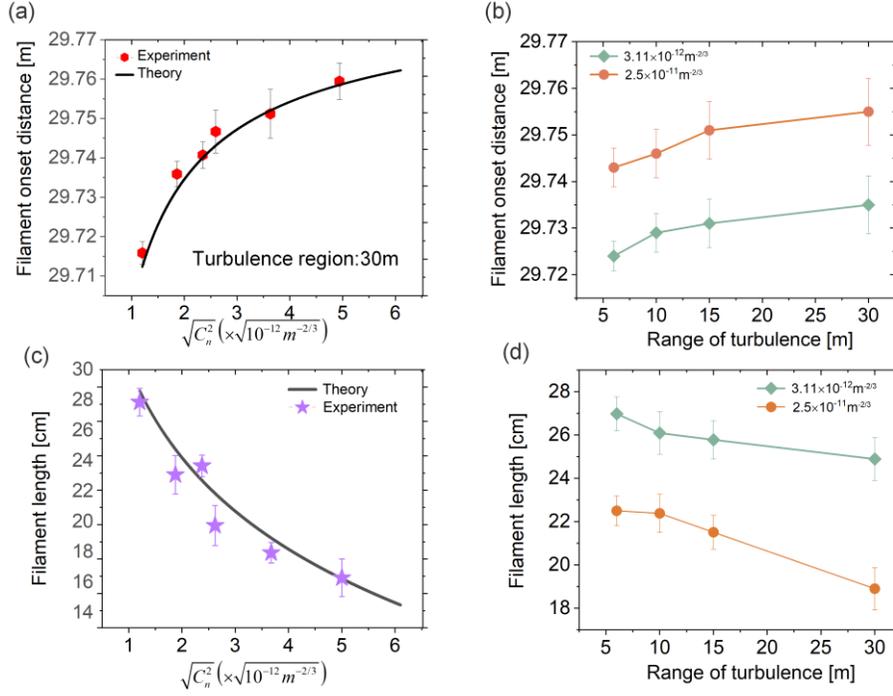

Fig. 2 (a)Onset position and (c)filament length of filaments as a function of turbulence intensity in 30 m regions; (b)Onset position and (d)filament length of filaments as a function of turbulence regions.

Nevertheless, we can see that the relationship between the onset position of the filament and the square root of the turbulent structure constant is not linear, which may be due to the fact that the formation of the filament is a nonlinear process and the onset position of the filament is related to the self-focusing threshold power of the beam collapse. Turbulence has the potential to influence the self-focusing threshold power of femtosecond laser filamentation. The relationship between the turbulent refractive index structure constant and the self-focusing threshold power has been studied theoretically, and air turbulence is predicted to increases the power threshold ($P_c$) for beam collapse[21], which can be described as $\frac{P_c}{P_{cr}}=1+\frac{3}{4}k_0^2 a^2 \left(\frac{aC}{2}\right)^{2/3}$, where $P_{cr}=\frac{3.77\lambda^2}{8\pi n_2 n_0}$ is the critical power threshold, $C=4.38 l_0^{-1/3} C_n^2 \left\{1-[1+17.5\frac{a^2}{l_0^2}]^{-1/6}\right\}$. Therefore, we can obtain the relationship between the self-focusing threshold power under the turbulent and the filament starting position. Consequently, the Air turbulence impacts the self-focusing distance of the filament, i.e., $Z_f = \dfrac{0.367 k_0 a^2}{\sqrt{(\sqrt{\dfrac{P}{P_c}}-0.852)^2 - 0.0219}}$. Additionally, the self-focal distance

($Z_f$) will be modified under different external focusing conditions, i.e., $Z_f' = \frac{Z_f f}{Z_f + f}$ [22]. Fig. 2(a) indicates that the experimental and theoretical results are in good agreement.

Then we explore the impact of turbulent action region on the filament start position under two experimental conditions: maximum $2.5 \times 10^{-11} \text{m}^{-2/3}$ and minimum $3.11 \times 10^{-12} \text{m}^{-2/3}$ turbulence intensity. As depicted in Fig. 2(b), the start position of filaments moves slightly away from the focusing lens as the turbulent action area increases. We suppose that the move is attributed to the cumulative effect of turbulence increasing as the turbulent action area expands. Based on the observations from Fig. 2(a) and Fig. 2(b), it is evident that when the turbulence intensity is $2.5 \times 10^{-11} \text{m}^{-2/3}$ and the turbulent action area extends to 30 m, the start position of filament undergoes a displacement of approximately 5 cm towards the direction of laser propagation, in contrast to its start position in the laboratory environment.

Furthermore, the variations of turbulence intensity and turbulence region influence the length of femtosecond laser filament is investigated. Hence, we derive an estimation for the length of femtosecond laser filaments in turbulent indicated as the black curve in Fig. 2c. Fig. 2(c) presents the filament lengths measured at various turbulence intensities, demonstrating a notable reduction in the length of the femtosecond laser filament under turbulent conditions. It is evident that the relationship between the filament length and the structure constant of the turbulent refractive index does not exhibit a linear decrease. It is worth noting that the change of the turbulence region does not alter the relationship between filament length and turbulence intensity (Fig. S2 in Supplementary Materials). The relationship between the turbulent refractive index structure constant and the length of filament has been studied theoretically, which has been reported that the length of the filament can be estimated using the following formula[23]:

$$L_f = L_r \times \left( \frac{1.1 \ln(0.81(P/P_c))}{1 + 0.3(f/L_r)^{-2.5}} \right) \quad (1)$$

where $L_r = k_0 a^2$. Fig. 2(c) indicates that the experimental and theoretical results are in good agreement.

We subsequently investigate the influence of turbulent action regions on the length of filaments under two experimental conditions: maximum and minimum turbulence intensity. Fig. 2(d) demonstrates that the increase of the turbulent action region leads to a further reduction in the length of the filament. Based on the findings from Fig. 2(c) and Fig. 2(d), when the turbulence intensity reaches $2.5 \times 10^{-11} \text{m}^{-2/3}$ and the turbulence region expands to 30 m, the length of filament measures 19 cm. Comparatively, the length is reduced by approximately 10 cm compared to the filament in the laboratory environment. Our estimation attributes this reduction primarily to the alteration of the filament formation threshold power in the presence of turbulence.

Air turbulence not only leads to an increase in the distance required for the chirp pulse to form the filament, but also causes an increased random drift in the center of the filament when the turbulence is occurs. Chin's group conducted experimental characterization of the drift of various distances from the center of the filament[24]. The experimental results revealed the absence of correlation between the displacements of the filament in the $x$ and $y$ directions. Additionally, it is proposed that the displacement of the filament center follows the Rayleigh distribution law. Here, our focus is directed towards investigating the influence of turbulence intensity and the region affected by turbulence on the center drift of filaments during the remote propagation of femtosecond laser pulses over a distance of 30 meters. Moreover, the pointing stability of optical filaments has been quantified by calculating the standard deviation of the beam center position recorded with Fig. 1(c). The beam center coordinates ($x_c, y_c$) are calculated according to [25]:

$$x_c = \sum_x \sum_y x \cdot S(x,y) \Big/ \sum_x \sum_y S(x,y) \tag{2a}$$

$$y_c = \sum_x \sum_y y \cdot S(x,y) \Big/ \sum_x \sum_y S(x,y) \tag{2b}$$

where $S$ indicates the intensity of the pixel.

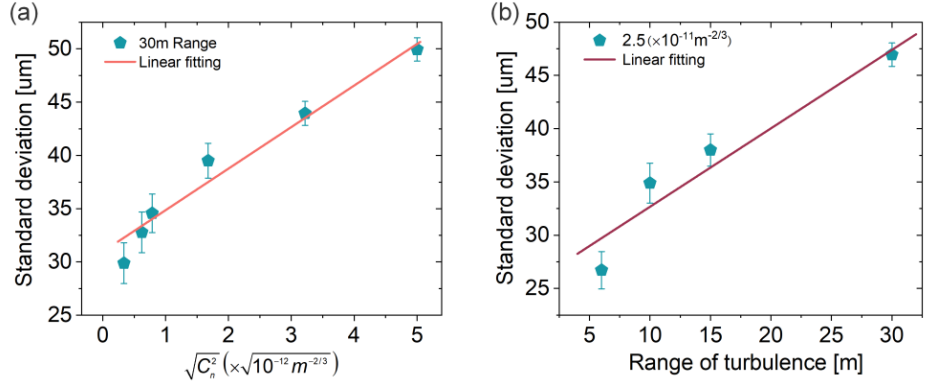

Fig. 3 Standard deviation of transverse displacement of filament depends on turbulence intensity in different regions.

Fig. 3(a) presents the observed drift of the standard deviation pertaining to the transverse displacement of the femtosecond laser filament as the turbulence intensity increases from $1.68 \times 10^{-12}\,\mathrm{m}^{-2/3}$ to $2.5 \times 10^{-11}\,\mathrm{m}^{-2/3}$. Notably, the enhanced turbulence intensity amplifies the drift of filament. Additionally, a linear relationship is observed between the recorded standard deviation and the square root of the turbulence intensity, demonstrating a satisfactory linear fit, which forms an empirical formula as follow:

$$\langle \delta \rangle = k \times \sqrt{C_n^2} \tag{3}$$

Importantly, it is worth emphasizing that the alteration of the turbulence region does not affect the relationship between standard deviation of transverse displacement of filament and turbulence intensity (Fig. S3 in Supplementary Materials).

We subsequently explore the impact of turbulent action regions on the standard deviation of transverse displacement of filaments under maximum turbulence intensity. Fig. 3(b) reveals that the expansion of the turbulent region leads to an increase in filament drift. Moreover, the standard deviation of filament drift exhibits a linear relationship with the enlargement of the turbulent region. This observation suggests that the cumulative effect on filament drift due to the turbulent region expansion follows a linear trend.

Next, we directed our attention to the influence of turbulence on the cross-sectional effective area of femtosecond laser filamentation. In Fig. 4(a), the diameter of the filament cross section exhibits a gradual increment as the turbulence intensity escalates from $1.68 \times 10^{-12}\,\mathrm{m}^{-2/3}$ to $2.5 \times 10^{-11}\,\mathrm{m}^{-2/3}$, considering a turbulent action region of 30 m. It is noteworthy that due to the non-standard circular shape of the filament cross section, we utilize the average of the sum of diameters in both the transverse and longitudinal directions as the effective diameter of the filament cross section. Furthermore, as the turbulence intensity increases, Fig. 4(a) demonstrates the increase of wavefront instability caused by atmospheric turbulence, resulting in the formation of multiple filaments within the femtosecond laser. This observation aligns with previously reported findings, corroborating the consistency of our results[26]. As can be seen in Fig. 4(a), the femtosecond laser filament has formed a multifilament when the turbulence intensity is $8.37 \times 10^{-12}\,\mathrm{m}^{-2/3}$.

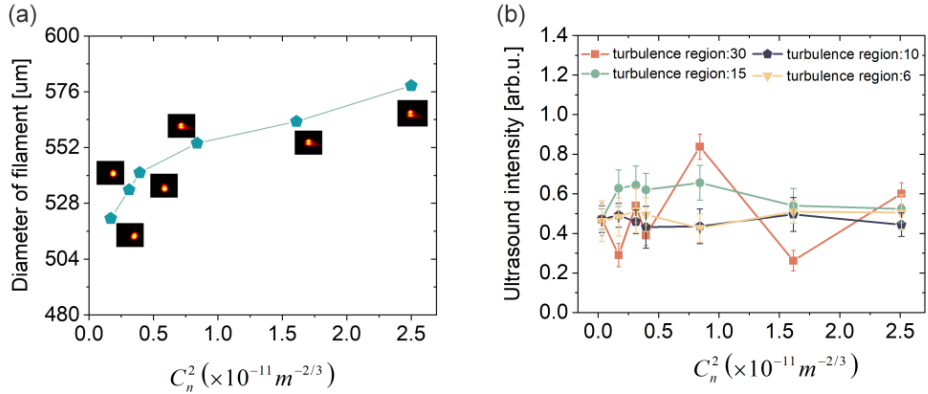

Fig. 4 Diameter of filament depends on turbulence intensity in 30 m regions

Lastly, we explored the impact of turbulence intensity on the intensity of the femtosecond laser filament. Employing the setup depicted in Fig. 1b, the radiated ultrasound signal intensity of the filament was recorded in various turbulence intensity environments, as depicted in Fig. 4(b). Fig. 4(b) demonstrates that alterations within the turbulent region, as well as variations in turbulence intensity, do not affect the intensity of the femtosecond laser filament. This phenomenon can primarily be attributed to the influence of turbulence on the formation threshold of the femtosecond laser filament, resulting in a corresponding alteration in the filament length. In the presence of turbulence, filaments formed exhibit a consistent intensity that remains unaffected by the turbulence.

## 4. Conclusion

In this paper, we have conducted experimental investigations on the impact of turbulence intensity and turbulence action region on various characteristics of femtosecond laser filaments. Specifically, the onset position, length, intensity, and cross-sectional diameter of these filaments were explored. By analyzing the ultrasonic signals emitted by the filaments, it is observed that increasing turbulence intensity and expanding turbulent action area cause the onset of filament to shift towards the direction of laser propagation, resulting in shorter filament length. Remarkably, these experimental findings align with theoretical calculations. Furthermore, it is observed that as turbulence intensity transitions from $1.68\times10^{-12}\,\mathrm{m}^{-2/3}$ to $2.5\times10^{-11}\,\mathrm{m}^{-2/3}$, the effective cross-sectional area of the femtosecond laser filament increases. Additionally, when turbulence intensity reaches at $8.37\times10^{-12}\,\mathrm{m}^{-2/3}$, multiple filaments are formed. Moreover, we have discovered a linear relationship between the standard deviation of the drift displacement of the femtosecond laser filament in its cross section and the square root of the turbulent intensity. Interestingly, it is determined that turbulence does not significantly affect the intensity of the femtosecond laser filament. Our findings provide valuable insights into the behavior of femtosecond laser propagation in complex environments, specifically atmospheric turbulence. These results contribute to the comprehension of long-range propagation mechanisms in complex atmospheric conditions. Furthermore, the results offer guidance for remote sensing and imaging applications aimed at detecting and analyzing multi-component pollutants and targets.

**Acknowledgments**

This work was supported by National Key Research and Development Program of China (2018YFB0504400).; Fundamental Research Funds for the Central Universities (63223052).

**Conflict of interest**

The authors declare no competing interests.